\documentstyle[Qqa4lart]{article}
\input Tcilatex

\begin{document}
\begin{center}
{\bf The Analytical Solution of Radiation Transfer Equation for a Layer of
Magnetized Plasma With Random Irregularities \\[20mm]} {\rm N.A.Zabotin} and 
{\rm A.G.Bronin }{\bf \\[20mm]} {\rm Rostov State University,
Rostov-on-Don,344090, Russia}
\end{center}

\begin{abstract}
The problem of radio wave reflection from an optically thick plane
monotonous layer of magnetized plasma is considered at present work. The
plasma electron density irregularities are described by spatial spectrum of
an arbitrary form. The small-angle scattering approximation \-in the
invariant ray coordinates is suggested for analytical investigation of the
radiation transfer equation. The approximated solution describing
spatial-and-angular distribution of radiation reflected from a plasma layer
is obtained. The obtained solution can be applied, for example, to the
ionospheric radio wave propagation.
\end{abstract}

{{{{\footnotesize {{{{\large {\rm {}}}}}} }}}}

\section{Introduction}

Basic goal of the present work consists in derivation of the transfer
equation solution describing spatial-and-angular distribution \ $P\left( 
\vec{\rho },\omega \right) ${} of radio radiation reflected from a plane
stratified layer of magnetized plasma with random irregularities.

{\footnotesize {{\large {\rm {}}}}}The radiation transfer equation (RTE) in
a randomly irregular magnetized plasma was obtained in the work [1] under
rather general initial assumptions. In particular, the medium average
properties were assumed smoothly varying both in space and in time. In the
work [2] the radiation energy balance (REB) equation describing radiation
transfer in a plane stratified layer of stationary plasma with random
irregularities has been deduced. The invariant ray coordinates, allowing one
to take into account by a natural way refraction of waves and to represent
the equation in the most simple form, were used there. In the work [3] it
was shown that the equation REB is a particular case of the radiation
transfer equation obtained in [1] and can be deduced from the latter by
means of transition to the invariant ray coordinates.Equation REB, thus,
allows one to investigate influence of multiple scattering in a plane
stratified plasma layer on the characteristics of radiation. In particular,
it enables one to determine the spatial-and-angular distribution of
radiation leaving the layer if the source directivity diagram and
irregularity spatial spectrum are known. A few effects which require of wave
amplitudes coherent summation for their description (for example, phenomenon
of enhanced backscattering) are excluded from consideration. However, the
multiple scattering effects are much stronger, as a rule. This is
particularly true for the ionospheric radio propagation.

The numerical methods of the transfer equation solving developed in the
theory of neutron transfer and in the atmospheric optics appear useless for
the equation REB analysis. They are adapted, basically, to the solution of
one-dimensional problems with isotropic scattering and plane incident wave.
In a case of magnetized plasma the presence of regular refraction,
aspect-sensitive character of scattering on anisometric irregularities and
high dimension of the equation REB (it contains two angular and two spatial
coordinates as independent variables) complicate construction of the
effective numerical algorithm for its solving. In this situation it is
expedient to solve the equation REB in two stages. The first stage consists
of obtaining of the approximated analytical solution allowing one to carry
out the qualitative analysis of its properties and to reveal of its
peculiarities. At the second stage the numerical estimation methods can be
applied to the obtained analytical solution, or methods of the numerical
solving of the initial equation taking into account the information obtained
at the first stage can be designed. Therefore the problem of obtaining of
the equation REB approximated analytical solutions is of interest.

{}We begin the present paper from a detailed exposition of the invariant ray
coordinates concept. Then possibility to use of the small-angle scattering
in the invariant coordinates approximation is discussed. Two modifications
of the REB equation solution are obtained. The analysis of the obtained
solutions concludes the paper.

\section{Invariant ray coordinates and the radiation energy balance equation{%
{{{{{{{{{{\protect\footnotesize {{{{{{{{{{{\protect\large {\rm \ }}}}}}}}}}}}%
}}}}}}}}}}}}

It is convenient to display graphically the electromagnetic wave propagation
in a plane-stratified plasma layer with the aid of the Poeverlein
construction [4,5]. We shall briefly describe it. Let the Cartesian system
of coordinates has axis $z$\ {} perpendicular and the plane $x0y$\ {}
parallel to the plasma layer. We shall name such coordinate system
``vertical''. It is assumed that the vector of the external magnetic field \
{}$\vec H$ is situated in the plane $z0y$\ {}. Module of radius-vector of
any point inside of the unit sphere with centrum in the coordinate origin
corresponds to the value of refractive index\ $n_i(v,\alpha )${}, where $i=1$%
\ {} relates to the extraordinary wave, $i=$\ {} relates to the ordinary
one, $v=\omega _e^2/\omega ^2$\ {}, $\omega _e^2$\ {} is the plasma
frequency, \ {}$\omega ^2$ is the frequency of a wave, \ $\alpha ${} is the
angle between radius-vector and magnetic field $\vec H$\ {}. The refractive
index surface corresponding to a fixed value of \ \ $v${} and to all
possible directions of the radius-vector represents a rotation body about an
axis parallel to vector \ $\vec H$\ {} (see fig. 1).

Convenience of the described construction (in fact, this is an example of
coordinate system in space of wave vectors \ $\vec k${}) is become evident
when drawing the wave trajectory: it is represented by a straight line,
parallel to the axis $z$\ . This is a consequence of the generalized Snell
law, which also requires of equality of the fall angle and exit angle
onto/from a layer ($\theta $\ ), and constantness of the wave vector azimuth
angle (\ $\varphi ${}). Note, that the crossing point of a wave trajectory
with the refractive index surface under given value of \ {}$v$ determines
current direction of the wave vector in a layer (it is anti-parallel to a
radius-vector) and current direction of the group speed vector (it coincides
with the normal to the refractive index surface). The projection of a wave
trajectory onto the plane \ $x0y$\ is a point which radius-vector has module 
$\sin \theta ${} and its angle with relation to axis \ $x$\ equals to $%
\varphi $\ {}. Thus, the coordinates \ {} define completely the whole ray
trajectory shape in a plane layer and outside of it and are, in this sense,
invariant on this trajectory.

Radiation of an arbitrary point source of electromagnetic waves within the
solid angle $\theta \div \theta +d\theta ;\varphi \div \varphi +d\varphi $\
corresponds to the energy flux in the \ $\vec k${}-space inside of a
cylindrical ray tube parallel to axis \ $z${} with cross section $\sin
\theta d(\sin \theta )d\varphi =\sin \theta \cos \theta d\theta d\varphi $\
. In case of regular (without random irregularities) plasma layer this
energy flux is conserved and completely determined by the source directivity
diagram:

\begin{equation}
\label{eq1}P(z;\theta ,\varphi ,\vec \rho )=P_0(\theta ,\,\varphi ,\vec \rho
)\,\,, 
\end{equation}
where $P$\ is energy flux density in the direction determined by angles $%
\theta ,\varphi $\ {} through the point $\vec \rho $\ {} on some base plane
situated outside of the layer parallel to it (in the ionosphere case it is
convenient to choose the Earth' surface as the base plane), \ $z$\ is
distance from the base plane (height in the ionosphere case). We shall
assume in the present paper that function $z(v)$\ {} is monotonous in the
region of wave propagation and reflection. If random irregularities are
absent and source of radiation is point, variable \ $\vec \rho $\ in (\ref
{eq1}) is superfluous, as the matter of fact, since unequivocal relation
between it and angles of arrival of a ray \ $\theta ,\varphi $\ {} exists.
When scattering is present the radiation energy redistributes over angular
variables \ {}$\theta ,\varphi $ and in space what is described by variable $%
\vec \rho $\ {}. The value of $P$\ {} satisfies in this case to the equation
of radiation energy balance [2,3]:{\footnotesize {\large {\rm \ }}} 
\begin{equation}
\label{eq2}
\begin{array}{l}
\frac d{dz}P(z,\theta ,\varphi ,\vec \rho )=\int \{-P(z;\theta ,\varphi
,\vec \rho )\sin \theta \cos \theta C^{-1}(z;\theta ,\varphi )\cdot \\ 
\cdot \sigma \left[ \alpha _0\left( \theta ,\varphi \right) ,\beta _0\left(
\theta ,\varphi \right) ;\alpha \left( \theta ^{\prime },\varphi ^{\prime
}\right) ,\beta \left( \theta ^{\prime },\varphi ^{\prime }\right) \right]
\sin \alpha (\theta ^{\prime },\varphi ^{\prime })\left| 
\frac{\partial (\alpha ,\beta )}{\partial (\theta ^{\prime },\varphi
^{\prime })}\right| + \\ +P\left[ z;\theta ^{\prime },\varphi ^{\prime
},\vec \rho -\vec \Phi (z;\theta ^{\prime },\varphi ^{\prime };\theta
,\varphi )\right] \sin \theta ^{\prime }\cos \theta ^{\prime
}C^{-1}(z;\theta ^{\prime },\varphi ^{\prime })\cdot \\ 
\cdot \sigma \left[ \alpha _0\left( \theta ^{\prime },\varphi ^{\prime
}\right) ,\beta _0\left( \theta ^{\prime },\varphi ^{\prime }\right) ;\alpha
\left( \theta ,\varphi \right) ,\beta \left( \theta ,\varphi \right) \right]
\sin \alpha (\theta ,\varphi )\left| \frac{\partial (\alpha ,\beta )}{%
\partial (\theta ,\varphi )}\right| \}d\theta ^{\prime }d\varphi ^{\prime } 
\end{array}
\end{equation}
$C(z;\theta ,\varphi )$\ is cosine of a ray trajectory inclination angle
corresponding to the invariant angles \ $\theta $\ and $\varphi ;\left|
\partial (\alpha ,\beta )/\partial (\theta ,\varphi \right| $ is Jacobean of
transition from angular coordinates $\theta ,\varphi $\ \ to the wave vector
polar and azimuth angles $\alpha $\ \ and \ $\beta $\ in the ``magnetic''
coordinate system (which axis $0z$\ \ is parallel to the magnetic field); \ 
$$
\left[ \alpha _0\left( \theta ,\varphi \right) ,\beta _0\left( \theta
,\varphi \right) ;\alpha \left( \theta ^{\prime },\varphi ^{\prime }\right)
,\beta \left( \theta ^{\prime }.\varphi ^{\prime }\right) \right] \equiv
\sigma \left[ \theta ,\varphi ;\theta ^{\prime },\varphi ^{\prime }\right] 
{\rm {\ }} 
$$
is scattering differential cross section describing intensity of the
scattered wave with wave vector coordinates $\alpha ,\beta $\ \ in magnetic
coordinate system (corresponding invariant coordinates are \ $\theta
^{\prime }$\ and \ $\varphi ^{\prime }$) which arises at interaction of the
wave with wave vector coordinates $\alpha _0,\beta _0$\ (invariant
coordinates $\theta ${\large {\rm {\ \ }}}and $\varphi $\ ) with
irregularities. Vector function $\vec \Phi (z;\theta ^{\prime },\varphi
^{\prime };\theta ,\varphi )$\ represents the displacement of the point of
arrival onto the base plane of a ray which has angular coordinates $\theta
^{\prime }$\ \ and $\varphi ^{\prime }$\ \ after scattering at level $z$\ \
with relation to the point of arrival of an incident ray with angular
coordinates $\theta ,\varphi $\ . It is essential that in a plane-stratified
medium the function \ $\vec \Phi $\ is determined only by smoothed layer
structure \ $v(z)$\ and does not depend on the scattering point horizontal
coordinate and also on coordinate $\vec \rho $\ \ of the incident and
scattered rays. Note also that ratio $\vec \Phi (z;\theta ,\varphi ;\theta
^{\prime },\varphi ^{\prime })=-\vec \Phi (z;\theta ^{\prime },\varphi
^{\prime };\theta ,\varphi )$\ \ takes place.

It is possible to check up that equation (2) satisfies to the energy
conservation law: when integrating over all possible for level $z$\ \ values
of $\theta ,\varphi $\ and $\,$all $\,\vec \rho $\ its right side turns into
zero. It is natural since in absence of true absorption the energy inside
the plasma layer does not collected.

\ Analyzing expression for the scattering differential cross section in a
magnetized plasma (see, for example, [6]), it is easy to be convinced that
the following symmetry ratio takes place:

\begin{equation}
\label{eq3}\sigma \left[ \theta ,\varphi ;\theta ^{\prime },\varphi ^{\prime
}\right] \,n^2\cos \vartheta _g^{\prime }=\sigma \left[ \theta ^{\prime
},\varphi ^{\prime };\theta ,\varphi \right] n^{\prime 2}\cos \vartheta _g 
\end{equation}
where $\vartheta _g$\ {} is angle between the wave vector and group speed
vector, $n$\ {} is refractive index. Using (\ref{eq3}) the equation (\ref
{eq2}) can be presented as follows:

\begin{equation}
\label{eq4}
\begin{array}{l}
{\rm \ }\frac d{dz}P(z,\vec{\rho },\theta ,\varphi )=\int Q(z;\theta
,\varphi ;\theta ^{\prime },\varphi ^{\prime }) \\ \left\{ P(z,\vec{\rho }-%
{\rm \ }\vec{\Phi (z};\theta ^{\prime },\varphi ^{\prime };\theta ,\varphi
),\theta ^{\prime },\varphi ^{\prime })-P(z,\vec{\rho },\theta ,\varphi
)\right\} d\theta ^{\prime }d\varphi ^{\prime } 
\end{array}
\end{equation}
{\rm {}}where $Q(z;\theta ,\varphi ;\theta \prime ,\varphi \prime )=\sigma
(\theta ,\varphi ;\theta ^{\prime },\varphi ^{\prime })C^{-1}(z,\theta
,\varphi )\sin \theta ^{\prime }\left| d\Omega _k^{\prime }/d\Omega ^{\prime
}\right| $\ , and quantity $\stackrel{\sim }{Q}(z;\theta ,\varphi ;\theta
^{\prime },\varphi ^{\prime })$\ $\equiv Q(z;\theta ,\varphi ;\theta
^{\prime },\varphi ^{\prime })\sin \theta \cos \theta ${} is symmetric with
relation to rearrangement of shaded and not shaded variables. The equation
REB in the form (\ref{eq4}) has the most compact and perfect appearance. It
is clear from physical reasons that (\ref{eq4}) has to have the unique
solution for given initial distribution $P_0(\theta ,\varphi ,\vec{\rho }).$%
\ \ The obtained equation can be directly used for numerical calculation of
the signal strength spatial distribution in presence of scattering. However,
as it was noted at introduction already, this approach leads to essential
difficulties. Subsequent sections describe the method of construction of the
energy balance equation approximated analytical solution.

\section{Small-angle scattering approximation in the invariant ray
coordinates}

Let us consider the auxiliary equation of the following kind, which differs
from (4) only by absence of the dash over variable\ $\omega $\ marked by
arrow:

\begin{equation}
\label{eq5}{\rm \ }\frac d{dz}P(z,\vec \rho ,\omega )=\int Q(z;\omega
;\omega ^{\prime })\left\{ P(z,\vec \rho +{\rm \ }\vec \Phi (z;\omega
;\omega ^{\prime }),\stackunder{\uparrow }{\omega })-P(z,\vec \rho ,\omega
)\right\} d\omega ^{\prime } 
\end{equation}
where designation $\omega =\left\{ \theta ,\varphi \right\} ,\,\,d\omega
=d\theta d\varphi ${} has been used for the sake of compactness. Equation (%
\ref{eq5}) can be easily solved analytically by means of Fourier
transformation over variable $\vec \rho $.\ \ The solution has the following
form:

\begin{equation}
\label{eq6}P(z,\vec{q},\omega )=P_0(\vec {q},\omega )S(z,0;\vec{q},\omega ), 
\end{equation}
where $P_0(\vec q,\omega )${} is the Fourier image of the radiation energy
flux density passing the layer in absence of scattering and the value of \ $%
S $\ is defined by the expression

\begin{equation}
\label{eq7}S(z_2,z_1,\vec{q},\omega )=\exp \left\{
\int_{z_1}^{z_2}dz^{\prime }\int d\omega ^{\prime }Q(z^{\prime };\omega
;\omega ^{\prime })\left[ \exp \left( i\vec{q}\vec {\Phi }(z^{\prime
};\omega ;\omega ^{\prime }),\stackunder{}{\omega })\right) -1\right]
\right\} 
\end{equation}
One should note that integration over \ $z${} in this and subsequent
formulae, in fact, corresponds to integration along the ray trajectory with
parameters $\theta ,\varphi $.\ The area of integration over $\omega
^{\prime }$\ {} includes rays which reflection level $h_r(\omega ^{\prime
})>z$\ .

{}Let us transform now equation (\ref{eq4}) by the following way:

\begin{equation}
\label{eq8}
\begin{array}{c}
\frac d{dz}P(z, 
\vec{\rho },\omega )=\int d\omega ^{\prime }Q(z;\omega ;\omega ^{\prime
})\left\{ P(z,\vec{\rho }+{\rm \ }\vec{\Phi }(z;\omega ;\omega ^{\prime
}),\omega )-P(z,\vec{\rho },\omega )\right\} + \\ +\int d\omega ^{\prime
}Q(z;\omega ;\omega ^{\prime })\left\{ P(z,\vec{\rho }+{\rm \ }\vec{\Phi }%
(z;\omega ;\omega ^{\prime }),\omega ^{\prime })-\,P(z,\vec{\rho }+{\rm \ }%
\vec {\Phi }(z;\omega ;\omega ^{\prime }),\omega )\right\} 
\end{array}
\end{equation}
Its solution will be looked for in the form

\begin{equation}
\label{eq9}P(z,\vec \rho ,\omega )=\stackrel{\sim }{P}(z,\vec \rho ,\omega
)+X(z,\vec \rho ,\omega ) 
\end{equation}
Thus, auxiliary equation (\ref{eq5}) allows to present the solution of the
equation (\ref{eq4}) in the form (\ref{eq9}). This is an exact
representation while some approximated expressions for quantities $\stackrel{%
\sim }{P}$\ \ and $X$ \ \ are not used.

{}By substituting of (\ref{eq9}) into the equation (\ref{eq4}) one can
obtain the following equation for the unknown function $X$\ :

\begin{equation}
\label{eq10}
\begin{array}{l}
\frac d{dz}X(z,\vec \rho ,\omega )=\int d\omega ^{\prime }Q(z;\omega ;\omega
^{\prime })\{ 
\stackrel{\sim }{P}(z,\vec \rho +{\rm \ }\vec \Phi (z;\omega ;\omega
^{\prime }),\omega ^{\prime })-\, \\ - 
\stackrel{\sim }{P}(z,\vec \rho +{\rm \ }\vec \Phi (z;\omega ;\omega
^{\prime }),\omega )\}+\int d\omega ^{\prime }Q(z;\omega ;\omega ^{\prime
})\{X(z,\vec \rho +{\rm \ }\vec \Phi (z;\omega ;\omega ^{\prime }),\omega
^{\prime })-\, \\ -\,X(z,\vec \rho ,\omega )\} 
\end{array}
\end{equation}
We shall assume now that the most probable distinction of angles $\omega
^{\prime }$\ \ and \ $\omega $\ is small. The heuristic basis for this
assumption is given by analysis of the Poeverlein construction (fig. 1). It
is easy to be convinced examining the Poeverlein construction that
scattering near the reflection level even for large angles in the wave
vectors space entails small changes of the invariant angles $\theta ,\varphi 
$.\ \ This is especially true for irregularities strongly stretched along
the magnetic field (in this case the edges of scattered waves wave vectors
form circles shown in fig. 1 as patterns A and B). One should note also that
the changes of invariant angles $\theta ,\varphi $\ {} are certainly small
if scattering with small change of a wave vector direction takes place. This
situation is typical for irregularity spectra, in which irregularities with
scales more than sounding wave length dominate. Thus, the small-angle
scattering approximation in the invariant coordinates has wider
applicability area than common small-angle scattering approximation.

{}Scattering with small changes of \ $\theta ,\varphi $\ entails small value
of \ $\left| \vec \Phi \right| $\ That follows directly both from sense of
this quantity and from the fact what$\left| \vec \Phi (z,\omega ,\omega
)\right| =0$\ . Let us make use of that to carry out expansion of quantity $%
X $\ {} at the right side of the equation (10) into the Taylor series with
small quantities $\omega ^{\prime }-\omega $ \ {}and \ $\left| \vec \Phi
\right| $\ .{} Note that making similar expansion of function $P$\ \ at the
initial equation (\ref{eq4}) would be incorrect since function P may not to
have property of continuity. For example, in case of a point source, $P_0$\
{} is a combination of \ $\delta $\ {}-functions. As it will be shown later,
the function $X$\ {} is expressed from $P_0$\ {} by means of repeated
integration and, hence, differentiability condition fulfills much easier for
it.

{}Leaving after expansion only small quantities of the first order, we
obtain the following equation in partial derivatives:

\begin{equation}
\label{eq11}\frac \partial {\partial z}X(z,\vec{\rho },\omega )-A_\omega
(z,\omega )\frac \partial {\partial \omega }X(z,\vec {\rho },\omega )+A_{%
\vec{\rho }}(z,\omega )\frac \partial {\partial \vec{\rho }}X(z,\vec {\rho
},\omega )=f({\rm \ }z,\vec{\rho },\omega ), 
\end{equation}
where

$$
A_\omega (z,\omega )=\int d\omega ^{\prime }Q(z;\omega ;\omega ^{\prime
})(\omega ^{\prime }-\omega ); 
$$

$$
A_{\vec{\rho }}(z,\omega )=\int d\omega ^{\prime }Q(z;\omega ;\omega
^{\prime })\vec \Phi (z,\omega ,\omega ^{\prime }); 
$$

$$
\begin{array}{l}
f( 
{\rm \ }z,\vec \rho ,\omega )=\int d\omega ^{\prime }Q(z;\omega ;\omega
^{\prime }) \\ \left\{ \stackrel{\sim }{P}(z,\vec \rho +{\rm \ }\vec \Phi
(z;\omega ;\omega ^{\prime }),\omega ^{\prime })-\,\stackrel{\sim }{P}%
(z,\vec \rho +{\rm \ }\vec \Phi (z;\omega ;\omega ^{\prime }),\omega
)\right\} 
\end{array}
$$
Here is the characteristic system for the equation (\ref{eq11}):

$$
\frac{dX}{dz}=f({\rm \ }z,\vec{\rho },\omega );\,\,\,\frac{d\vec{\rho }}{dz}%
=A_{\vec{\rho }}(z,\omega );\,\,\,\,\frac{d\omega }{dz}=-A_\omega (z,\omega
), 
$$
and initial conditions for it at $z=0$:\ 

$$
X=0;\,\vec\rho ^{\prime }=\vec{\rho };\,\omega =\omega _0. 
$$
It is necessary to emphasize the distinction between quantity $\vec\rho
^{\prime }$\ {}, which is a function of $z$\ , and invariant variable $\vec{%
\rho }$.

\ Solving the characteristic system we obtain:

$$
\,\omega =\omega (z,\omega _0),\,\,\,\,\,\,\,\,\,\vec \rho ^{\prime }=\vec{%
\rho }-\int_z^{z_0}dz^{\prime }A_{\vec{\rho }}(z^{\prime },\omega (z^{\prime
},\omega _0))\,. 
$$
where $z_0$\ {} is $z$\ {}-coordinate of the base plane. It follows that

\begin{equation}
\label{eq12}X(z_0,\vec \rho ,\omega )=\int_0^{z_0}dz^{\prime }f\left\{ {\rm %
\ }z^{\prime },\vec \rho -\int_{z^{\prime }}^{z_0}dz^{\prime \prime }A_{\vec
\rho }\left[ z^{\prime \prime },\omega (z^{\prime \prime },\omega _0)\right]
,\omega (z^{\prime },\omega _0)\right\} 
\end{equation}
Generally, expression (\ref{eq12}) gives the exact solution of the equation (%
\ref{eq11}). However, since we are already within the framework of the
invariant coordinate small-angle scattering approximation which assumes
small value of $A_\omega (z,\omega )$\ {}, it is possible to simplify the
problem a little. Assuming $A_\omega \cong 0$\ and omitting index 0 at
invariant coordinates $\omega $\ {}, we are coming to the following
approximate representation for function $X\,$:

\begin{equation}
\label{eq13}
\begin{array}{l}
X(z_0, 
\vec{\rho },\omega )=\int_0^{z_0}dz^{\prime }\{\stackrel{\sim }{P}\left[ z,%
\vec{\rho }+{\rm \ }\vec{\Phi }(z^{\prime };\omega ;\omega ^{\prime })+\vec
{D}(z_0,z^{\prime },\omega ),\omega ^{\prime }\right] - \\ \stackrel{\sim }{P%
}\left[ z^{\prime },\vec{\rho }+{\rm \ }\vec{\Phi }(z;\omega ;\omega
^{\prime })+\vec {D}(z_0,z^{\prime },\omega ),\omega \right] \} 
\end{array}
\end{equation}
\ where $\vec{D}(z_2,z_1,\omega )=\int_{z_1}^{z_2}\int d\omega ^{\prime
}Q(z;\omega ;\omega ^{\prime })\vec \Phi (z,\omega ,\omega ^{\prime })$\ .

\ Thus, in the invariant coordinate small-angle scattering approximation the
solution of the equation REB (\ref{eq4}) is represented as a sum of two
terms (see (\ref{eq9})), the first of which is

\begin{equation}
\label{eq14}
\begin{array}{l}
\stackrel{\sim }{P}(z,\vec{\rho },\omega )=\frac 1{\left( 2\pi \right)
^2}\,\int d^2qP_0(\vec q,\omega )\cdot \\ \cdot {\rm \exp \left\{ i\vec
q\vec \rho +\int_0^{z_0}dz^{\prime }\int d\omega ^{\prime }Q(z^{\prime
};\omega ;\omega ^{\prime })\left[ \exp \left( i\vec {q}\vec{\Phi }%
(z^{\prime };\omega ;\omega ^{\prime }),\stackunder{}{\omega })\right)
-1\right] \right\} } 
\end{array}
\end{equation}
{}where \ $\frac 1{\left( 2\pi \right) ^2}\int d^2qP_0(\vec q,\omega )\exp
(i\vec q\vec \rho )=P_0(\vec \rho ,\omega )${}, and the second one is given
by expression (\ref{eq13}).

\ The solution can be presented in the most simple form if one uses again
the smallness of quantity $\left| \vec \Phi \right| $\ {} and expands the
second exponent in the formula (\ref{eq14}) into a series. Leaving after
expansion only small quantities of the first order, one can obtain:

\begin{equation}
\label{eq15}
\begin{array}{l}
P(z, 
\vec{\rho },\omega )\cong P\text{{\rm $_0\left[ \vec \rho +\vec{D}%
(z_0,0,\omega ),\omega \right] \,+$}}\int_0^{z_0}dz^{\prime }\int d\omega
^{\prime }Q(z^{\prime };\omega ;\omega ^{\prime })\cdot \\ \cdot \{ 
\stackrel{\sim }{P}\left[ z,\vec{\rho }+{\rm \ }\vec {\Phi }(z^{\prime
};\omega ;\omega ^{\prime })+\vec {D}(z_0,z^{\prime },\omega )+\vec{D}%
(z^{\prime },0,\omega ^{\prime }),\omega ^{\prime }\right] -\, \\ -\stackrel{%
\sim }{P}\left[ z^{\prime },\vec{\rho }+{\rm \ }\vec{\Phi }(z;\omega ;\omega
^{\prime })+\vec{D}(z_0,0,\omega ),\omega \right] \} 
\end{array}
\end{equation}
The last operation is the more precise the faster value of \ {}$P_0(\vec
q,\omega )$ decreases under $\left| \vec q\right| \to \infty ${}. \ The
solution of the radiation energy balance equation obtained in the present
section in the form (\ref{eq9}), (\ref{eq14}), (\ref{eq13}), or in the form (%
\ref{eq15}), expresses the spatial-and-angular distribution of radiation
intensity passing layer of plasma with scattering through the
spatial-and-angular distribution of the incident radiation, that is, in
essence, through the source directivity diagram.

\section{Alternative approach in solving the REB equation}

{}The REB equation solving method stated in the previous section is based on
representation of quantity $P(z,\vec{\rho },\omega )$\ \ {} as a sum of the
singular part $\stackrel{\sim }{P}(z,\vec{\rho },\omega )$\ {} and the
regular one $X(z_0,\vec{\rho },\omega )$\ {}. Regularity of the $X(z_0,\vec{%
\rho },\omega )$\ {} has allowed one to use the expansion into the Taylor
series over variables $\vec{\rho }$\ {} and $\omega $\ {} at the equation (%
\ref{eq10}) right side and to transform the integral-differential equation (%
\ref{eq10}) into the first order partial derivative differential equation (%
\ref{eq11}).

{}However, the stated approach is not the only possible. The REB equation
can be transformed right away using Fourier-representation of the function $%
P(z,\vec{\rho },\omega )$\ {}$:$\ 

\begin{equation}
\label{eq16}{\rm P(z,\vec{\rho },\omega )=\frac 1{\left( 2\pi \right)
^2}\int d^2qP(z,\vec q,\omega )\exp (i\vec q\vec \rho )} 
\end{equation}
Substitution of (\ref{eq16}) into (\ref{eq4}) gives the following equation
for quantity $P(z,\vec{\rho },\omega )$\ {}:

\begin{equation}
\label{eq17}{\rm \ }\frac d{dz}P(z,\vec q,\omega )=\int d\omega ^{\prime
}Q(z;\omega ;\omega ^{\prime })\left\{ P(z,\vec q,\omega ^{\prime })\exp
\left( i\vec{q}\vec{\Phi }(z;\omega ;\omega ^{\prime })\right) -P(z,\vec{q}%
,\omega )\right\} 
\end{equation}
The quantity $P(z,\vec q,\omega )$\ \ {} is a differentiable function even
when $P(z,\vec{\rho },\omega )$\ {} has some peculiarities. Therefore, in
the invariant coordinate small-angle scattering approximation it is possible
to use the following expansion:

\begin{equation}
\label{eq18}P(z,\vec q,\omega ^{\prime })\cong P(z,\vec q,\omega )+\frac{%
\partial P(z,\vec q,\omega )}{\partial \omega }(\omega ^{\prime }-\omega
)\,. 
\end{equation}
Substituting (\ref{eq18}) in (\ref{eq17}) we obtain the partial derivative
differential equation

\begin{equation}
\label{eq19}\frac \partial {\partial z}P(z,\vec q,\omega )-\stackrel{\sim }{A%
}(z,\vec q,\omega )\frac \partial {\partial \omega }P(z,\vec q,\omega
)-P(z,\vec q,\omega )\stackrel{\sim }{S}(z,\vec q,\omega )=0\,, 
\end{equation}
where

{} 
$$
{\rm \stackrel{\sim }{S}(z,\vec q,\omega )=\int d\omega ^{\prime
}Q(z^{\prime };\omega ;\omega ^{\prime })\left[ \exp \left( i\vec{q}\vec{%
\Phi }(z^{\prime };\omega ;\omega ^{\prime }),\stackunder{}{\omega })\right)
-1\right] } 
$$

$$
{\rm \ }\stackrel{\sim }{A}(z,\vec q,\omega )={\rm \ \int d\omega ^{\prime
}Q(z^{\prime };\omega ;\omega ^{\prime })\exp \left( i\vec {q}\vec{\Phi }%
(z^{\prime };\omega ;\omega ^{\prime }),\stackunder{}{\omega })\right)
(\omega ^{\prime }-\omega )\ }. 
$$
The characteristic system

\begin{equation}
\label{eq20}\frac{d\omega }{dz}=-\stackrel{\sim }{A}(z,\vec q,\omega
),\,\,\,\,\,\,\,\frac{dP}{dz}=\stackrel{\sim }{S}(z,\vec q,\omega )P(z,\vec
q,\omega ) 
\end{equation}
with initial conditions \ $P=P_0(\vec q,\omega ),\omega =\omega _0$\ {}
at\thinspace $z=0$\ \ {} has the following solution:

\begin{equation}
\label{eq21}{\rm P(z,\vec{\rho },\omega )=\frac 1{\left( 2\pi \right)
^2}\int d^2qP_0(\vec q,\omega _0)\exp \left\{ i\vec q\vec \rho
+\int_0^zdz^{\prime }\stackrel{\sim }{S}\left[ z^{\prime },\vec{q},\omega
(z^{\prime },\vec q,\omega _0\right] \right\} } 
\end{equation}
This solution of the REB equation turns into the expression (\ref{eq14}) for 
$\stackrel{\sim }{P}$\ {} when $\stackrel{\sim }{A}(z,\vec q,\omega
)\longrightarrow 0$\ {}. But the latter limit transition corresponds to the
invariant coordinate small-angle scattering approximation used in the
previous section under derivation of (\ref{eq13}) and subsequent
expressions. Let us note, however, that in (\ref{eq21}), in contrast with (%
\ref{eq9}), any additional terms do not appear. It allows one to assume that
in used approximation the ratio

\begin{equation}
\label{eq22}{\rm X(z,\vec{\rho },\omega )\ll P(z,\vec{\rho },\omega )} 
\end{equation}
is fulfilled. Additional arguments to the benefit of this assumption will be
presented in the following section.

\section{Analysis of the solution of the REB equation}

\ {}We shall show, first of all, that the obtained solution satisfies to the
energy conservation law. For this purpose it is necessary to carry out
integration of the left and right sides of (\ref{eq15}) over $\omega $\ \
and $\vec \rho $\ \ multiplied them previously by \ $\sin \theta \cos \theta 
${}. The area of integration over angles is defined by the condition that
both wave \ $\omega $\ {} and wave \ $\omega ^{\prime }$\ {} achieve the
same level \ $z${} (since at level $z$\ {} their mutual scattering occurs).
To satisfy this condition one should add factors $\Theta \left[ h_r(\omega
)-z\right] $\ {}and $\Theta \left[ h_r(\omega ^{\prime })-z\right] $\ {}to
the integrand expression, where $\Theta (x)$\ {} is Heviside step function, $%
h_r(\omega )$\ \ {} is the maximum height which can be reached by a ray with
parameters $\theta ,\varphi $\ {}. Now integration can be expanded over all
possible values of angles, i.e., over interval \ {}$0\div \pi /2$\ for $%
\theta $\ {} and over interval $0\div 2\pi $\ {} for $\varphi $\ {}. Then, (%
\ref{eq15}) becomes

$$
\begin{array}{l}
\int P(\omega )\sin \theta \cos \theta d\omega =\int P_0(\omega )\sin \theta
\cos \theta d\omega + \\ 
\int_0^zdz^{\prime }\int d\omega \int d\omega ^{\prime }{\rm \Theta \left[
h_r(\omega )-z^{\prime }\right] {}\Theta \left[ h_r(\omega ^{\prime
})-z^{\prime }\right] }\stackrel{\sim }{Q}(z^{\prime };\omega ,\omega
^{\prime })\left[ P_0(\omega ^{\prime })-P_0(\omega )\right] {\rm \ } 
\end{array}
$$
where \ $P(\omega ),P_0(\omega )$\ {} is a result of integration of $P(z_0,%
\vec{\rho },\omega )$\ {} and $P_0(\vec {\rho },\omega )$\ {}
correspondingly over variable $\vec {\rho }$\ {}.

{}Due to antisymmetry of the integrand expression with relation to
rearrangement of shaded and not shaded variables, the last term in (\ref
{eq23}) is equal to zero. Thus, equation (\ref{eq23}) reduces to 
\begin{equation}
\label{eq24}\int P(z_0,\vec \rho ,\omega )\sin \theta \cos \theta d\omega
d^2\rho =\int P_0(\vec \rho ,\omega )\sin \theta \cos \theta d\omega d^2\rho 
\end{equation}
expressing the energy conservation law: the radiation energy full flux
\-through the plane \ {} remains constant regardless of scattering, as it
should be in case of real (dissipative) absorption absence. It is not
difficult to check that parity (\ref{eq24}) is valid for the exact solution
in the form (\ref{eq9}) and also for the solution in the form (\ref{eq21}).

{}With relation to the solution in the form (9) the carried out discussion
discovers one curious peculiarity. It appears that the radiation energy
complete flux through the base plane is determined by the first term $\left( 
\stackrel{\sim }{P}\right) $\ {}. The second one $(X)$\ \ gives zero
contribution to the energy complete flux.

{}Let us investigate in more detail the structure of quantity $X(z,\vec{\rho 
},\omega )$\ {} in the invariant coordinate small-angle scattering
approximation. Proceeding to the Fourier-representation in the expression (%
\ref{eq13}) produces

$$
\begin{array}{l}
X(z_0, 
\vec{q},\omega )=\int_0^{z_0}dz^{\prime }\int d\omega ^{\prime }Q(z^{\prime
};\omega ,\omega ^{\prime })\left[ \stackrel{\sim }{P}(z^{\prime },\vec{q}%
,\omega ^{\prime })-\stackrel{\sim }{P}(z^{\prime },\vec{q},\omega )\right]
\\ \exp \left\{ i\vec{q}\left[ \vec{\Phi }(z;\omega ;\omega ^{\prime })+\vec{%
D}(z_0,0,\omega )\right] \right\} 
\end{array}
$$
Employing regularity of function \ $\stackrel{\sim }{P}(z,\vec{q},\omega )$,
the last expression can be written as

$$
X(z_0,\vec{q},\omega )=\int_0^{z_0}dz^{\prime }\frac{\partial \stackrel{\sim 
}{P}(z,\vec{q},\omega )}{\partial \omega }\stackrel{\sim }{A}(z^{\prime },%
\vec{q},\omega )\exp \left[ i\vec{q}\vec{D}(z_0,z^{\prime },\omega )\right] 
$$
where quantity \ {}$\stackrel{\sim }{A}(z,\vec{q},\omega ) $ is defined by (%
\ref{eq19}). Thus, it becomes evident that limit transition $\stackrel{\sim 
}{A}(z,\vec{q},\omega )\longrightarrow 0$\ {} entails also $X(z_0,\vec{\rho }%
,\omega )\longrightarrow 0$\ \ {}. This property has been established in
section 4 with the aid of comparison of two variants of the REB equation
solution. Now we see that its presence is determined by structure of
quantity\ $X(z_0,\vec{\rho },\omega )$.

{}Results of the present section give the weighty ground to believe that the
radiation spatial-and-angular distribution is determined basically by the
first term in the solution (\ref{eq9}). The second term represents the
amendment to the solution which can be neglected in the invariant coordinate
small-angle scattering approximation. This statement validity can be checked
under detailed research of properties of the obtained REB equation
approximated solutions by numerical methods.

\section{Conclusion}

In the present work the heuristic basis for use of the invariant coordinate
small-angle scattering approximation is considered under solving of the RTE
for a magnetized plasma layer. Within the framework of this approximation
two versions of the analytical solution have been obtained. They describe
spatial-and-angular distribution of radiation reflected from a monotonous
plasma layer with small-scale irregularities.

{}The final physical conclusions about influence of the multiple scattering
effects in a layer of plasma on the spatial-and-angular characteristics of
radiation are possible on the basis of detailed numerical research of the
obtained solutions. Such research is a subject of other our works.

Acknowledgments. The work was carried out under support of Russian Basic
Research Foundation (grants No. 94-02-03337 and No. 96-02-18499).

\section{References}

1. A.G. Bronin, N.A. Zabotin, Sov. Phys. JETP 75(4), 633 (1992).

2. N.A. Zabotin, Izvestiya Vysshich Uchebnykh Zavedenii, Radiofizika, 36,
1075 (1993), in russian.

3. A.G. Bronin, N.A. Zabotin, Izvestiya Vysshich Uchebnykh Zavedenii,
Radiofizika, 36, 1163 (1993), in russian.

4. V.L.Ginzburg, Propagation of Electromagnetic Waves in Plasma, Moscow,
''Nauka'', 1967, in russian.

5. Budden K.G. Radio waves in the ionosphere. - Cambridge: University Press,
1961.

6. Electrodinamics of Plasma, edited by A.I.Akhiezer, Moscow, ''Nauka'',
1974, in russian.

\end{document}